\documentclass[final,5p,times,authoryear]{elsarticle}

\usepackage{newtxtext,newtxmath}
\usepackage{gensymb}

\usepackage{amssymb}
\usepackage{amsmath}
\usepackage[T1]{fontenc}
\usepackage{graphicx}
\usepackage{multirow}
\usepackage[T2A,T1]{fontenc}
\usepackage[utf8]{inputenc}
\usepackage[english]{babel}
\usepackage[colorlinks=true,linkcolor=blue,citecolor=blue,urlcolor=blue]{hyperref}
\usepackage{siunitx}
\usepackage{microtype}
\def\aj{Astron. J.}
\def\apj{Astrophys. J.}

\def\apjs{Astrophys. J. Suppl. Ser.}

\def\aap{Astron Astrophys.}

\def\mnras{MNRAS}

\usepackage{longtable}

\def\arcsec{\hbox{$^{\prime\prime}$}}
\def\arcmin{\hbox{$^{\prime}$}}

\def\desi{{DESI LIS}}
\def\ps1{{Pan-STARRS1}}
\def\doubleline{\vskip 3pt\hrule \vskip 1.5pt \hrule \vskip 5pt}

\journal{High Energy Astrophysics}

\begin{document}

\begin{frontmatter}

\title{Spectroscopic galaxy redshifts in the Peanut cluster -- a massive nearly head-on cluster merger shortly after pericenter passage}
%\title{Spectroscopic galaxy redshifts in the Peanut cluster -- a massive and peculiar Bullet cluster–like merger}
%\title{Measurements of redshifts and velocity dispersions of galaxies in the massive cluster SRGe\,J023820.8$+$200556 (Peanut).}

\author[1]{Igor Zaznobin\corref{cor1}}
\ead{zaznobin@cosmos.ru}
\cortext[cor1]{Corresponding author}

\author[1,2]{Natalya Lyskova}
\author[2,3]{Ilfan Bikmaev}
\author[1]{Rodion Burenin}
\author[4]{Arina Arshinova}
\author[1,5]{Eugene Churazov}
\author[4,6]{Sergey Dodonov}
\author[1,5]{Marat Gilfanov}
\author[1,5]{Ildar Khabibullin}
\author[2,3]{Irek Khamitov}
\author[4,6]{Sergey Kotov}
\author[1,4]{Alexey Moiseev}
\author[1]{Sergey Sazonov}
\author[1,5]{Rashid Sunyaev}
\author[2,3]{Mikhail Suslikov}
\author[4]{Roman Uklein}

\affiliation[1]{organization={Space Research Institute (IKI), Russian Academy of Sciences},
            city={Moscow},
            postcode={117997},
            country={Russia}}
\affiliation[2]{organization={Kazan Federal University (KFU)},
            city={Kazan},
            postcode={420008},
            country={Russia}}        
\affiliation[3]{organization={Tatarstan Academy of Sciences},
            city={Kazan},
            postcode={420011},
            country={Russia}}
\affiliation[4]{organization={Special Astrophysical Observatory (SAO), Russian Academy of Sciences},
            city={Nizhny Arkhyz},
            postcode={369167},
            country={Russia}}  
\affiliation[5]{organization={Max Planck Institute for Astrophysics (MPA)},
            city={Garching},
            postcode={85748},
            country={Germany}} 
\affiliation[6]{organization={Institute of Applied Astronomy (IAA), Russian Academy of Sciences},
            city={Saint Petersburg},
            postcode={191187},
            country={Russia}}
            
\begin{abstract}

The Peanut cluster (SRGe\,J023820.8$+$200556, SRGe\,CL0238.3+2005, $z_{\rm spec} = 0.42$) has recently emerged as a candidate for a rare, massive merger, potentially analogous to the Bullet cluster. We present the results of optical identification and spectroscopic redshift measurements for 31 galaxies in the Peanut cluster, including 26 new redshifts obtained with the 6-m telescope BTA (Big Telescope Alt-azimuthal) at SAO RAS between October 2024 and January 2025. The derived distribution of line-of-sight velocities reveals the possible presence of two subclusters with a line-of-sight velocity difference of $\sim$2000~km/s. However, statistical tests and the Dressler-Schectman test show that the hypothesis that the observed velocity distribution can be described by a normal distribution for a single cluster cannot be ruled out, and the evidence for the existence of two gravitationally bound substructures remains ambiguous. Assuming a single cluster with the normal velocity distribution, the estimated galaxy velocity dispersion is $\sigma_{los} = 1455 \pm 83$ km/s, corresponding to the total cluster mass of $M_{200}\simeq2\times10^{15}\,M_{\odot}$ based on the mass–velocity dispersion scaling relation. In either scenario -- a single extremely massive cluster or an ongoing merger -- the Peanut cluster appears to be a very rare and peculiar object, comparable to such extreme systems as the Bullet cluster (1E\,0657-56) or El~Gordo (ACT-CL\,J0102-4915).

\end{abstract}


\end{frontmatter}

%%%%%%%%%%%%%%%%%%%%%%%%%%%%%%%%%%%%%%%%%%%%%%%%%%

%%%%%%%%%%%%%%%%% BODY OF PAPER %%%%%%%%%%%%%%%%%%

\section{Introduction}

The all-sky survey conducted by the eROSITA X-ray telescope \citep{ero} aboard the Spektr-Roentgen-Gamma space observatory \citep[SRG,][]{srg} has enabled the construction of a large sample of galaxy clusters. This sample includes both morphologically regular and merging clusters. Studies of merging galaxy clusters are of particular interest, for example, for testing models of self-interacting dark matter \citep[see, e.g., the review by][]{2025RvMP...97d5004A}. Spectroscopic redshift measurements of galaxies in merging clusters can be used to identify individual gravitationally bound components and constrain their properties \citep[e.g.,][]{1988AJ.....95..985D, 2013MNRAS.430.3453B, 2023A&A...669A.147B}.

In June 2020, we initiated an observational program aimed at the optical identification and spectroscopic redshift measurements of the most massive galaxy clusters detected in the SRG/eROSITA all-sky survey. Observations were carried out using the 6m BTA telescope of the Special Astrophysical Observatory of the Russian Academy of Sciences (SAO RAS), the 1.6m AZT-33IK telescope of the Sayan Solar Observatory of the Institute of Solar-Terrestrial Physics SB RAS, and the 1.5m RTT-150 telescope of the T\"UB\.ITAK Observatory. As a result of this effort, by the end of 2022 we reported redshift measurements for several massive galaxy clusters in the Eastern Galactic hemisphere \citep{zazn21lh,br21,br22}, including the cluster SRGe\,J023820.8$+$200556 (the Peanut cluster). Spectroscopic redshift measurements for 216 galaxy clusters from this program were published in our subsequent work the following year \citep{cl216}.

In \cite{cl216} spectroscopic observations of the Peanut cluster were obtained with the SCORPIO-2 spectrograph \citep{scorpio11} on the BTA telescope during the night of 16–17 October 2020. For these observations, a single slit position with a zero degree position angle was selected to obtain spectra of the two brightest cluster galaxies. The measured  line-of-sight velocity difference between these galaxies was found to be 2000~km/s, which significantly  exceeds the expected velocity dispersion value\footnote{For the Peanut cluster, $M_{200}\simeq 1.4\times 10^{15} M_{\odot}$ \citep{Peanut}, and according to the mass–velocity dispersion scaling relation from \cite{2008ApJ...672..122E}, this mass corresponds to $\sigma_{\rm los} \simeq 1200$ km/s} for a cluster with a mass of $M_{200}\sim 10^{15} M_{\odot}$.

Following the analysis of detailed high-resolution Chandra data of the Peanut cluster, it was found that the cluster may be in a merger phase \citep{Peanut}. The X-ray image of the cluster reveals two bright substructures separated by a region of diminished surface brightness. Moreover, the peaks of the galaxy spatial distribution in the optical image do not coincide with the regions of enhanced X-ray emission. Thus, a spatial offset is observed between the hot X-ray gas and the collisionless galaxies. A classic example of such a spatial offset is the Bullet cluster, whose observations have provided constraints on the self-interaction cross-section of dark matter particles \citep{2008ApJ...679.1173R,2017MNRAS.465..569R}.

Thus, a comparison of the X-ray and optical images suggests a merger scenario. This interpretation is further supported by the line-of-sight velocity between the two cluster galaxies whose redshifts were measured with BTA in October 2020. The system is therefore most likely undergoing a merger and is observed close to pericenter passage. Therefore, a decision was made to obtain more extensive information on the line-of-sight velocities of the cluster galaxies, sufficient for a statistical analysis. To this end, we carried out observations with several optical telescopes, the results of which are presented in this work. 

The structure of the paper is as follows. Observations and data reduction are described in Sections 2 and 3, the analysis of the redshift distribution is presented in Section 4, and the conclusions are provided in Section 5. We adopt a $\Lambda$ cold dark matter ($\Lambda$CDM) cosmology with $\Omega_{\rm M}=0.3$, $\Omega_{\Lambda}=0.7$, and $H_0 = 70\ \mathrm{km\ s^{-1}\ Mpc^{-1}}$. At the cluster redshift $z=0.42$, 1 arcmin corresponds to 332 kpc. Throughout this paper, we use $M_{200}$ and $M_{500}$ to denote the masses enclosed within $r_{200}$ and $r_{500}$, where the mean density equals 200 and 500 times the critical density of the Universe at the cluster redshift.

\section{Additional observations}

The first additional observations of the Peanut cluster were carried out in summer 2024 with the RTT-150 telescope. These observations enabled us to measure spectroscopic redshifts for three additional bright cluster galaxies, and the results were reported in \cite{Peanut}. The new measurements further supported the hypothesis that the cluster is undergoing an active merger process.

By early September 2024, no galaxies in the Peanut cluster field brighter than $r=20$ mag remained without spectroscopic redshift measurements. We therefore continued the observational campaign with the 6-m BTA telescope using the SCORPIO-1 \citep{scorpio05} and SCORPIO-2 spectrographs \citep{scorpio11}.

\subsection{Object Selection}

Targets for subsequent spectroscopic observations with BTA were selected through identification of the red sequence of galaxies in the color–magnitude diagram \citep{deVac,RS}. For this purpose, we used photometric data for galaxies in the \emph{r} and \emph{z} bands of the Sloan filter system, taken from the DESI LIS survey \citep{desi}. Objects classified as extended sources in the DESI LIS survey and located within 2\arcmin\ from the center of X-ray emission (RA = 02:38:20.8, DEC = +20:05:56) were plotted on the color–magnitude diagram. As discussed in \cite{Peanut}, the adopted X-ray center does not correspond to the surface brightness peak, but instead represents  an approximate  centroid of the surface brightness distribution on scales of 1\arcmin–3\arcmin. For the five galaxies with spectroscopic redshifts reported in our previous work \citep{Peanut} the $r-z$ color and the $r$-band magnitude was approximated by a linear fit. The resulting fit shows no significant dependence of color on $r$-band magnitude; the mean color of these galaxies (with its standard deviation) is $(r-z)_{\rm Peanut} = 1.045 \pm 0.020$.

\begin{figure}
  \centering
    \includegraphics[width=0.97\columnwidth]{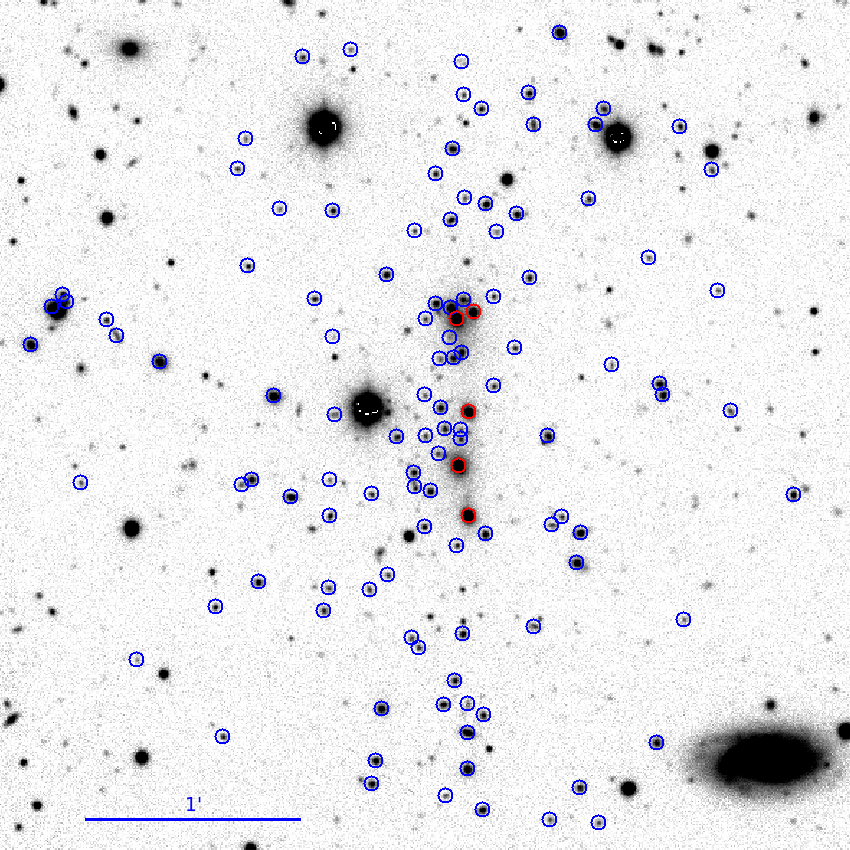}
    \caption{Galaxies in the Peanut cluster field on a $z$-band image from the \desi\ survey. Red circles mark the five galaxies with spectroscopic redshifts reported in \cite{Peanut}. Blue circles indicate 115 galaxies brighter than $r < 22$ within 2\arcmin\ of the X-ray source center that are possible cluster members.}
  \label{fig:RS_galaxies}
\end{figure}

For target selection, we manually increased the red-sequence width to  0.2 magnitudes, i.e., $(r-z)_{RS} \approx 1.045 \pm 0.200$. The color–magnitude diagram shows that this region contains a large number of galaxies brighter than $r=22$ magnitudes, compared to the region outside it. We therefore assume that this region predominantly contains galaxies belonging to the Peanut cluster, forming its red sequence in the color–magnitude diagram. Accordingly, for subsequent spectroscopic observations we selected galaxies in the central region of the cluster with $r\le22$ and $0.845 \leq r-z \leq 1.245$. Within a radius of 2\arcmin\ from the X-ray source center, 115 galaxies brighter than $r=22$ without measured spectroscopic redshifts satisfy this selection criterion (see Figure~\ref{fig:RS_galaxies}).

It is not feasible to measure spectroscopic redshifts for such a large number of galaxies with BTA using the long-slit mode within less than one year, given the observational time allocated to our group. We therefore limited the observational program to a small number of spectrograph slit positions. %When planning the observational program, 
The spectrograph slit positions were selected manually to maximize the number of bright red-sequence galaxies in the central cluster region captured within each slit. In October 2024, we selected nine slit positions for subsequent observations with BTA.

\subsection{Observations with BTA}

\begin{figure*}
  \centering
  \begin{minipage}{0.48\textwidth}
    \centering
    October 2020
    \end{minipage}
    \hfill
    \begin{minipage}{0.48\textwidth}
    \centering
    October 2024
    \end{minipage}
    \vspace{2cm}
  \includegraphics[width=0.98\columnwidth]{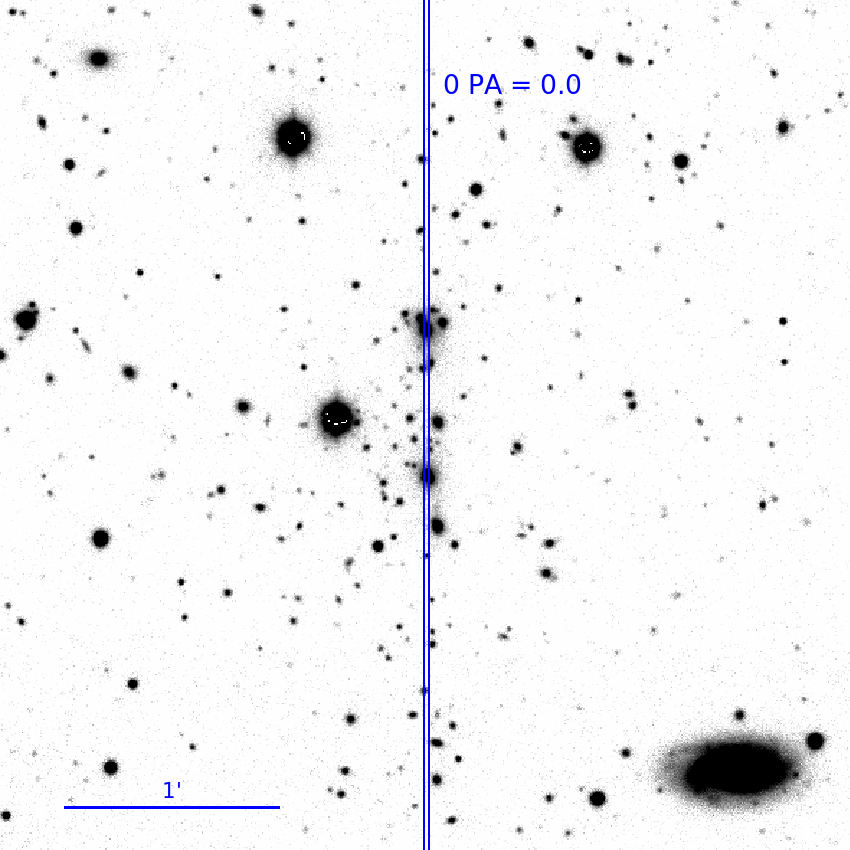}~ \includegraphics[width=0.98\columnwidth]{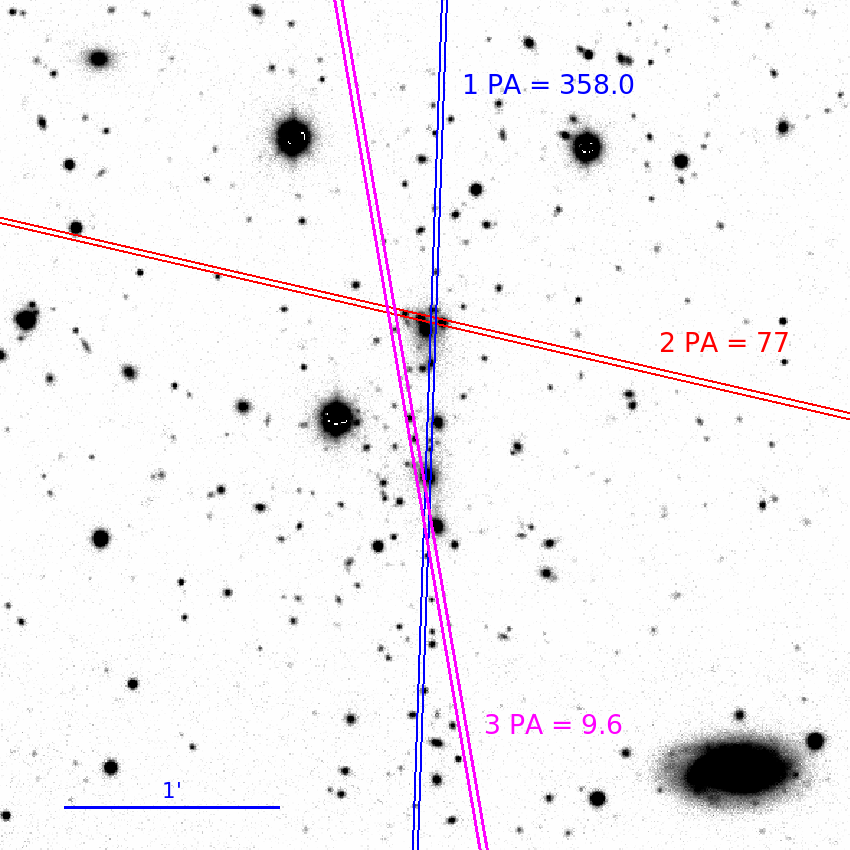}
    \begin{minipage}{0.48\textwidth}
    \centering
    December 2024
    \end{minipage}
    \hfill
    \begin{minipage}{0.48\textwidth}
    \centering
    January 2025
    \end{minipage}
  \includegraphics[width=0.98\columnwidth]{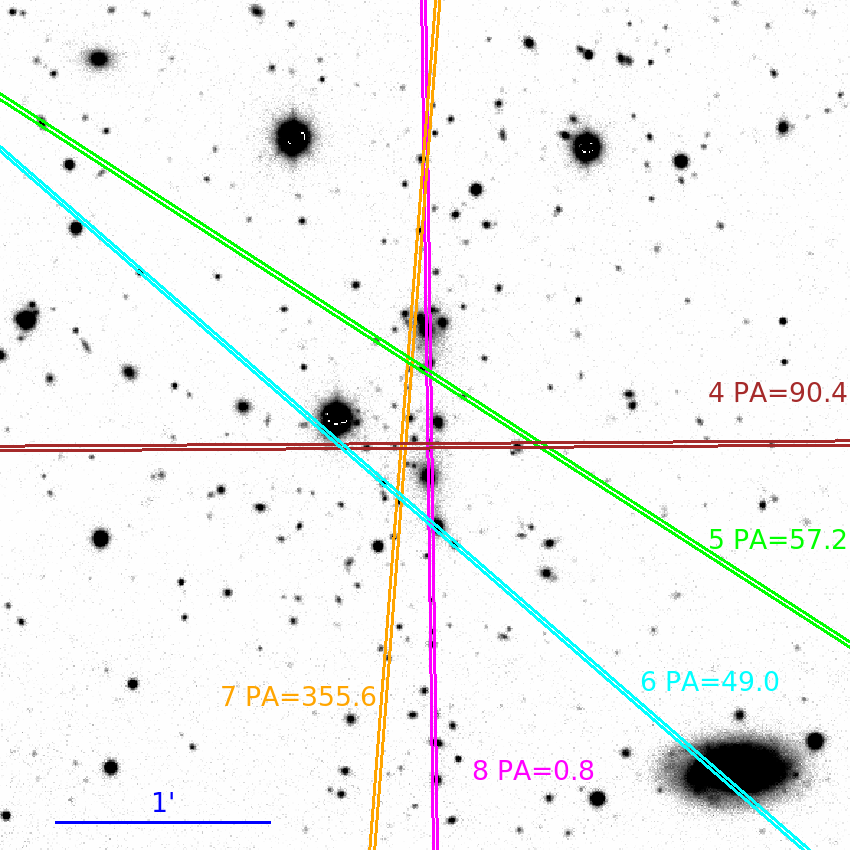}~    \includegraphics[width=0.98\columnwidth]{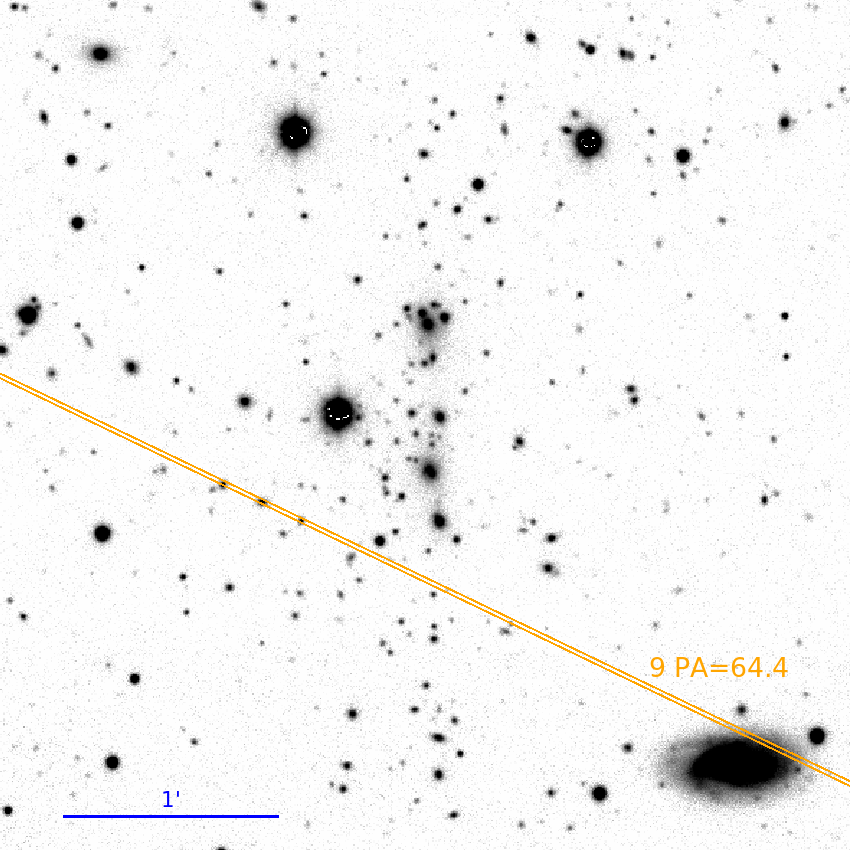}
  \caption{Finding charts graphically showing the outlines of the spectrograph slits with their labels and position angles indicated. The images are taken from the \desi\ survey in $z$-band.
    }
  \label{fig:slits1-9}
\end{figure*}

Given the cluster redshift of $z \simeq 0.42$, we selected the VPHG550G grism for observations with SCORPIO-1 and the VPHG940@600 grism for SCORPIO-2 as the most suitable configurations for obtaining the spectra. More detailed characteristics of the adopted spectral modes are given in Table~\ref{tab:grisms}. Observations for all nine proposed slit positions were carried out during dark moonless time in October and December 2024 and in January 2025. Table~\ref{tab:slits} provides detailed information on the observations performed, including conditional slit position numbers (positions), their central coordinates and position angles, as well as observation dates, image quality, exposure time, and number of exposures. Position 0 denotes the slit position used in October 2020, while the remaining positions are labeled from 1 to 9. For position 1, observations were conducted over two consecutive nights, since the initial total exposure of 1 hour was insufficient for a reliable redshift measurement, necessitating additional observations the following night. A graphical representation of positions 1–9 in the sky is shown in Figure~\ref{fig:slits1-9}.

\begin{table}
  \caption{Characteristics of the spectral modes used.}
  \renewcommand{\arraystretch}{1.0}
  \renewcommand{\tabcolsep}{0.1cm}
  \centering
  \footnotesize
  \begin{tabular}{lccccc}
    \noalign{\vskip 3pt\hrule\vskip 5pt}
    Grating & Spectral range & FWHM & Dispersion & Slit & Slit\\
    & \AA & \AA & \AA/pix & width, \arcsec & length, \arcmin\\
    \noalign{\vskip 3pt\hrule\vskip 5pt}
    VPHG500G & 3500--7500 & 13.6 & 2.1 & 1.2 & 6\\
    VPHG940@600 & 3500--8500 & 10.4 & 1.16 & 1.5 & 6\\
    \noalign{\vskip 3pt\hrule\vskip 5pt}
    \label{tab:grisms}
    \end{tabular}
\end{table}

\begin{table}
  \caption{BTA observation log}
  \renewcommand{\arraystretch}{1.0}
  \renewcommand{\tabcolsep}{0.05cm}
  \centering
  \footnotesize
  \begin{tabular}{cccccrcr}
    \noalign{\vskip 3pt\hrule\vskip 5pt}
    %Pos. & RA & Dec & Date & Device & PA & Seeing, \arcsec & Exp., sec.\\
    Pos. & RA & Dec & Date & Device & PA & Seeing & Exp.\\
     & & & & & Deg. & \arcsec & sec.\\
    \noalign{\vskip 3pt\hrule\vskip 5pt}
    0 & 02:38:20.7 & +20:05:52 & 2020-10-16 & SCORPIO-2 & 0.0   & 2.3 & 3$\times$600\\
    1 & 02:38:20.6 & +20:05:57 & 2024-10-04 & SCORPIO-1 & 358.0 & 1.2 & 4$\times$900\\ %Moiseev, Arshinova
      &            &           & 2024-10-05 & SCORPIO-1 & 358.0 & 1.5 & 5$\times$900\\ %Moiseev, Arshinova
    2 & 02:38:20.7 & +20:06:22 & 2024-10-04 & SCORPIO-1 & 77.0  & 1.3 & 7$\times$900\\ %Moiseev, Arshinova
    3 & 02:38:21.1 & +20:06:04 & 2024-10-05 & SCORPIO-1 & 9.6   & 1.5 & 6$\times$900\\ %Moiseev, Arshinova
    4 & 02:38:20.9 & +20:05:46 & 2024-12-01 & SCORPIO-1 & 90.4  & 1.5 & 8$\times$900\\ %Kotov, Arshinova
    5 & 02:38:20.4 & +20:06:04 & 2024-12-02 & SCORPIO-1 & 57.2  & 1.0 & 8$\times$900\\ %Kotov, Arshinova
    6 & 02:38:21.1 & +20:05:31 & 2024-12-02 & SCORPIO-1 & 49.0  & 1.0 & 10$\times$900\\ %Kotov, Arshinova
    7 & 02:38:21.0 & +20:06:02 & 2024-12-02 & SCORPIO-1 & 355.6 & 1.2 & 12$\times$900\\ %Kotov, Arshinova
    8 & 02:38:20.6 & +20:05:53 & 2024-12-03 & SCORPIO-1 & 0.8   & 1.2 & 12$\times$900\\ %Kotov, Arshinova
    9 & 02:38:23.9 & +20:05:29 & 2025-01-28 & SCORPIO-2 & 64.4 & 2.0 & 9$\times$900\\ %Dodonov
    \noalign{\vskip 3pt\hrule\vskip 5pt}
    \label{tab:slits}
    \end{tabular}
\end{table}

\section{Observational results}

When processing FITS images obtained during the observations, we performed extraction and reduction of all spectra present in the two-dimensional spectroscopic images, including those of faint galaxies. Spectroscopic image reduction was performed in a standard manner using the IRAF software package\footnote{https://iraf-community.github.io/} and custom software of the group. Redshifts were measured by comparing the galaxy spectra with a single stellar population (SSP) template with 7~Gyr age and metallicity $Z = 0.02$ \citep{bc03}. The galaxy redshift was determined as the local minimum of the $\chi^2$ statistic ($\chi^2_{min}$) obtained by comparing the observed spectra with the template. The error in measuring the redshift was defined as half the difference in redshifts at the $\chi^2_{min}+4$ level.

\begin{figure*}
  \centering
    \includegraphics[width=0.98\columnwidth]{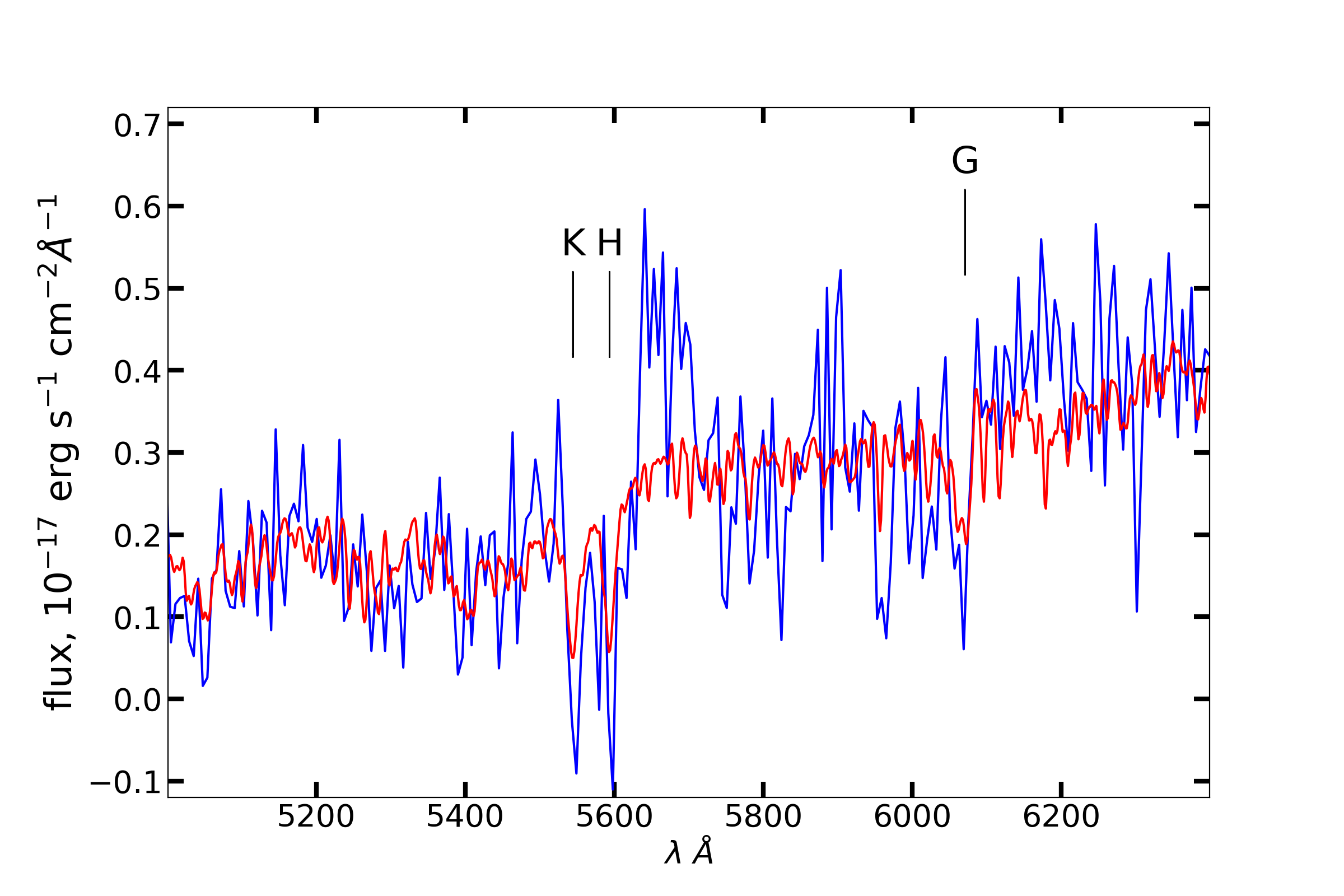}
    \includegraphics[width=0.98\columnwidth]{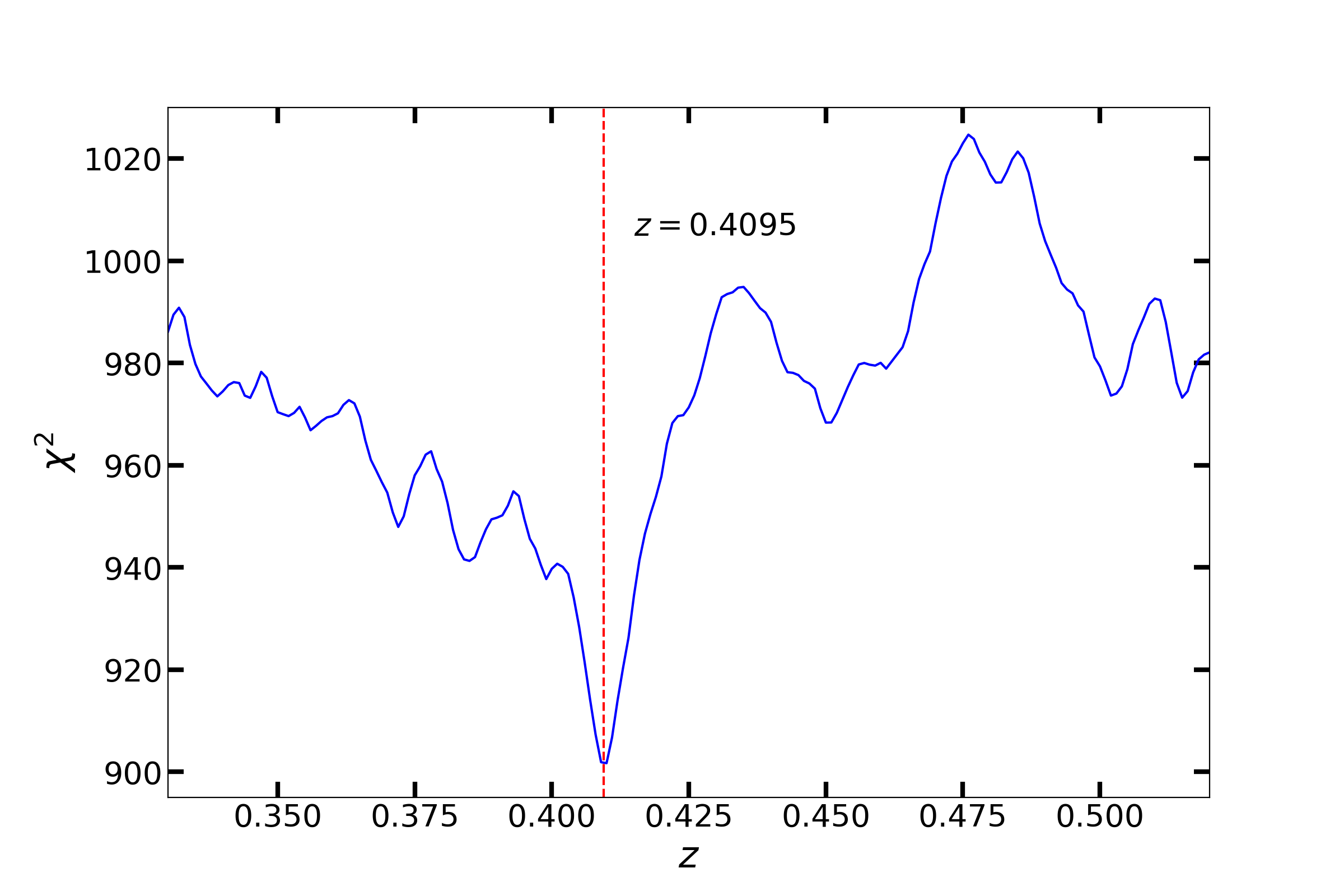}
    \includegraphics[width=0.98\columnwidth]{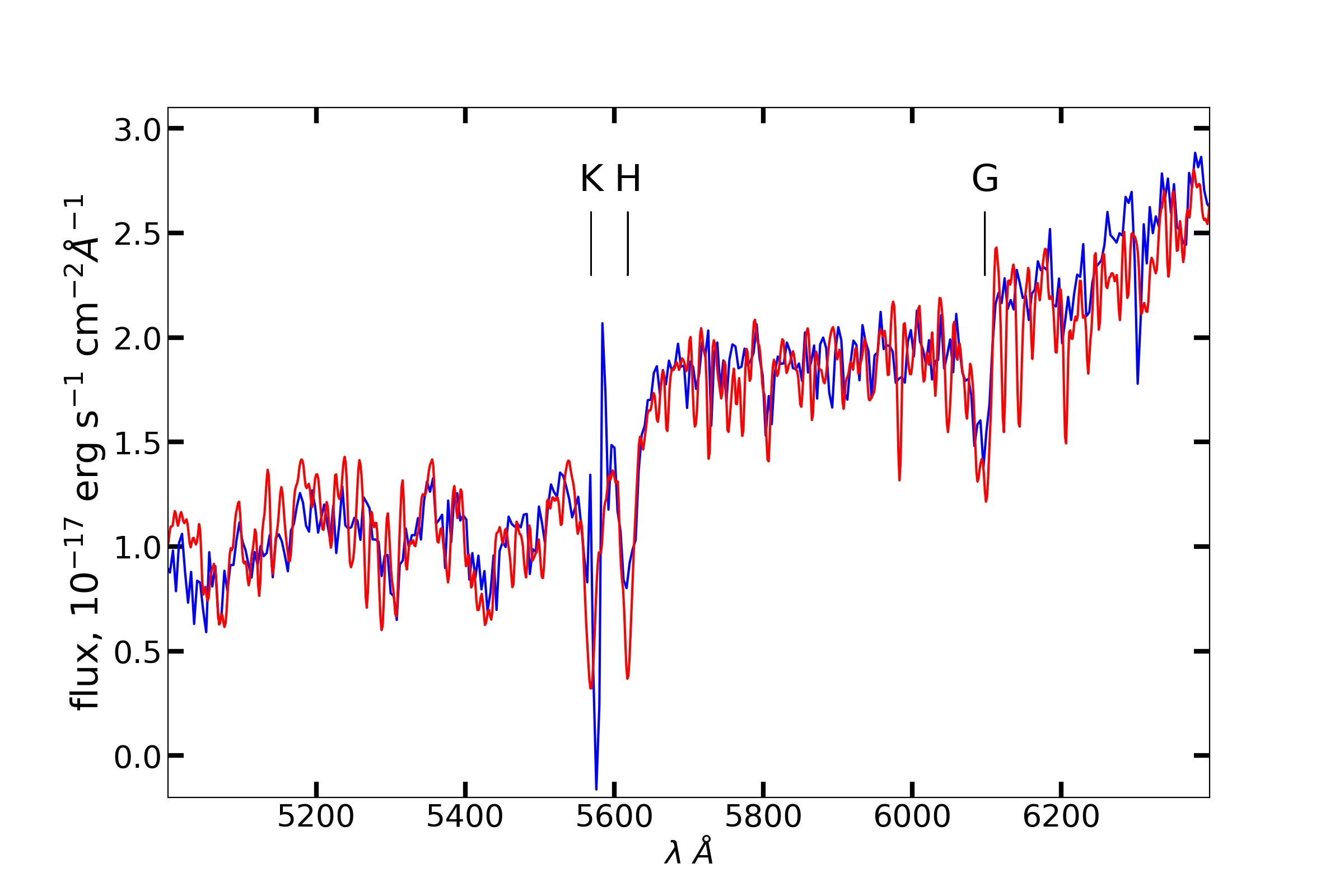}
    \includegraphics[width=0.98\columnwidth]{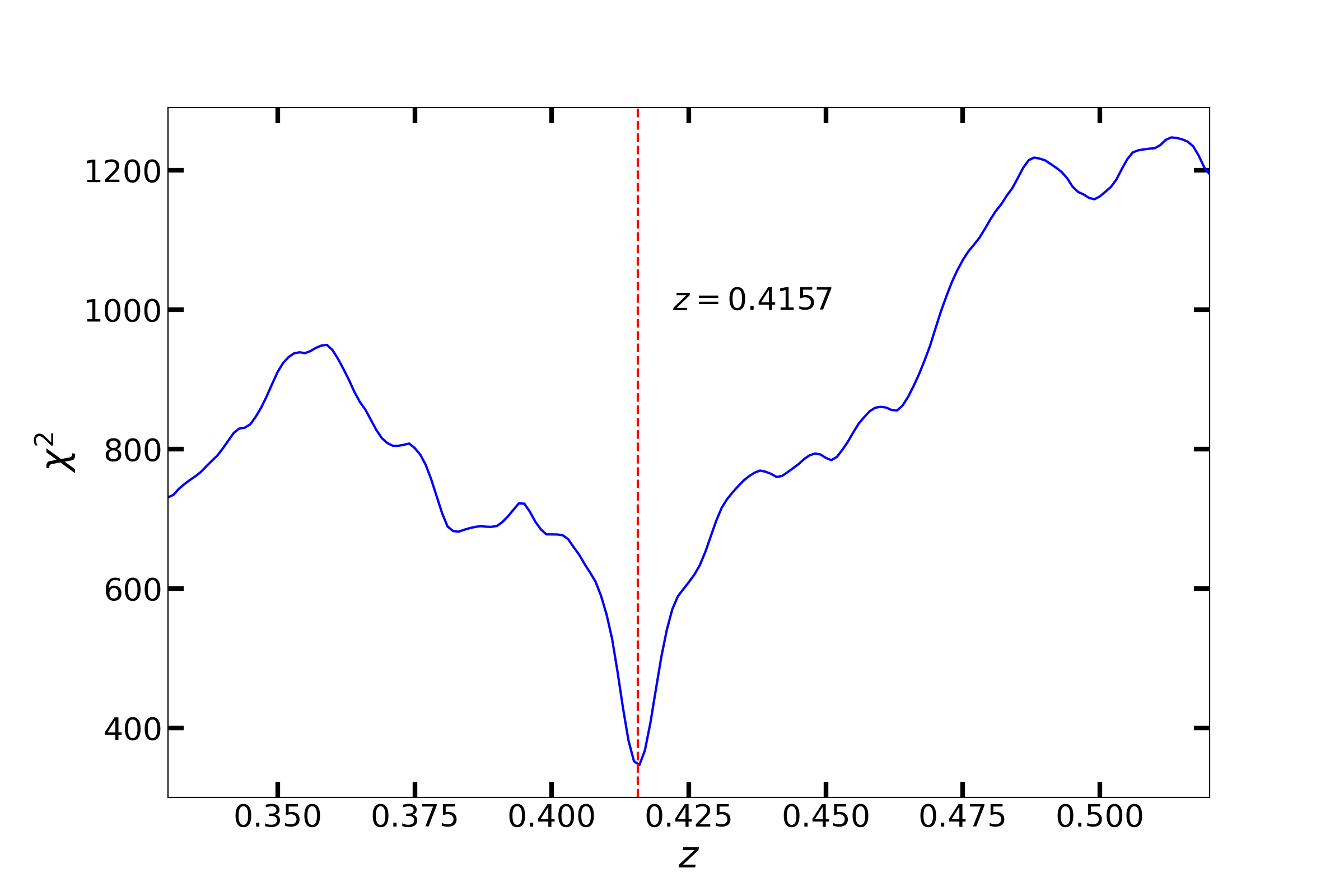}
    \includegraphics[width=0.98\columnwidth]{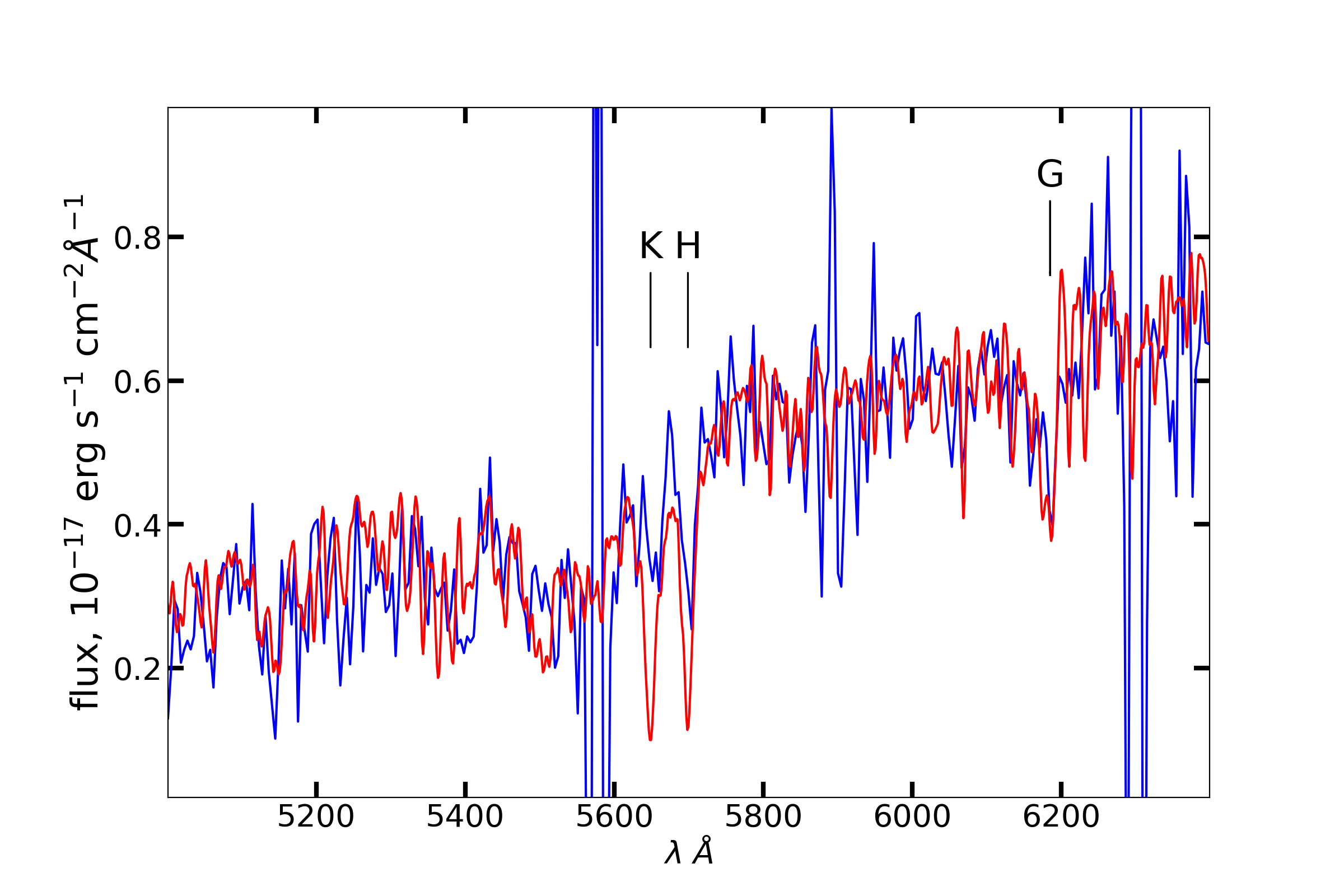}
    \includegraphics[width=0.98\columnwidth]{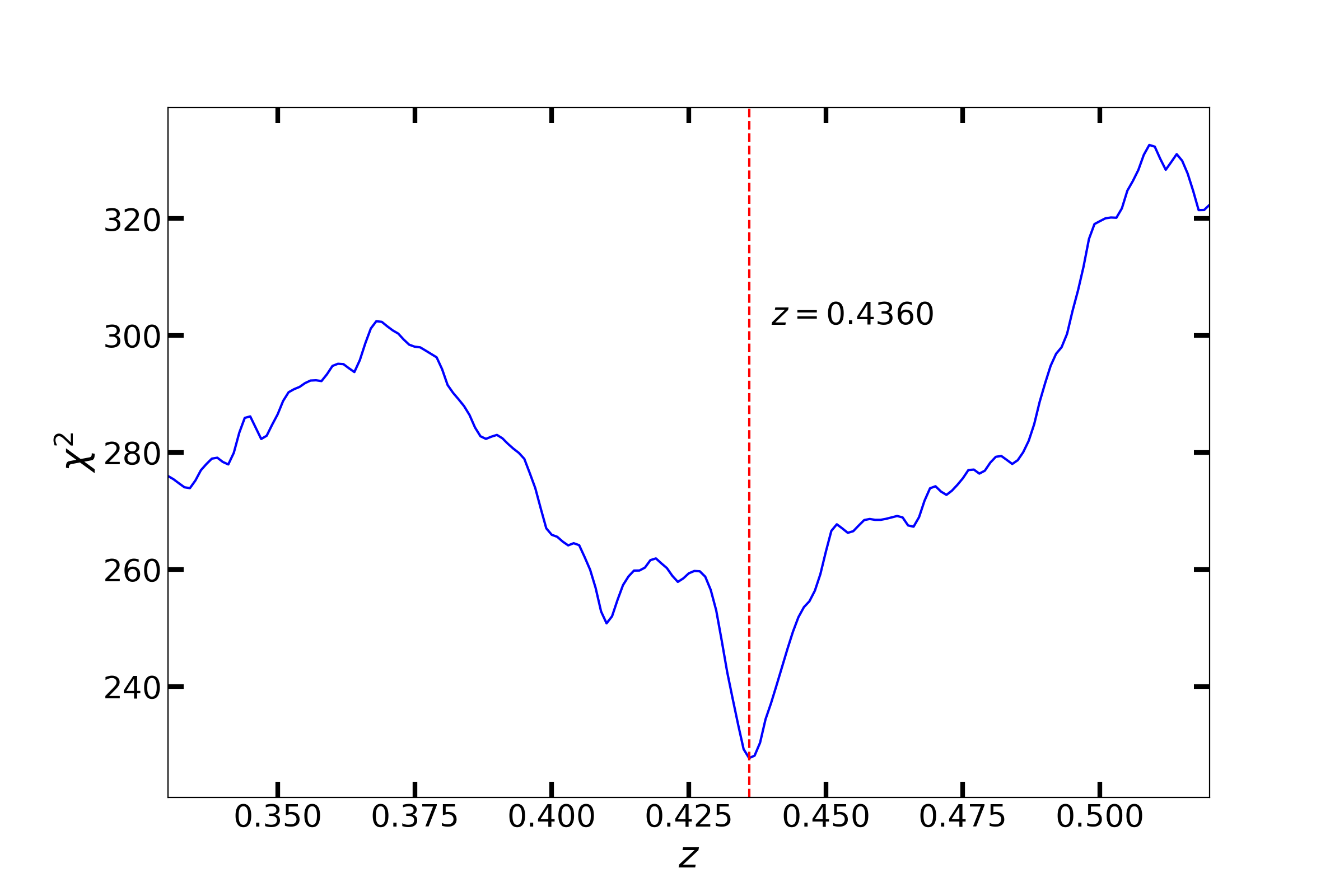}
    \caption{Examples of galaxy spectra. Left: observed cluster galaxy spectra (blue) with selected spectral features indicated, together with template spectra shifted to the corresponding galaxy redshifts (red). Right: $\chi^2$ distributions obtained from comparison of the observed spectra with the templates.}
  \label{fig:spec}
\end{figure*}

Examples of galaxy spectra are shown in Figure~\ref{fig:spec}. The left panels display the galaxy spectra, while the right panels show their corresponding $\chi^2$ distributions obtained from comparison with the template. The upper and lower rows show spectra of cluster galaxies with the lowest and highest redshifts, respectively, while the middle row presents the spectrum of the brightest cluster galaxy.

As a result, spectroscopic redshifts were measured for 26 galaxies, increasing the total number obtained by our group in the Peanut cluster field to 31. All measured redshifts lie within a narrow range of $z_{\rm spec}=0.4095$–0.4360, consistent with membership in the Peanut cluster at $z \simeq 0.42$.

\begin{table}
  \caption{Spectroscopic redshifts of galaxies}
  \renewcommand{\arraystretch}{1.0}
  \renewcommand{\tabcolsep}{0.13cm}
  \centering
  \footnotesize
  \begin{tabular}{cccclll}
    \noalign{\doubleline}
    RA & Dec & $r$ & $r-z$ & $z_{spec}$ & $z_{spec}^{err}$ & Pos.\\
    \noalign{\vskip 3pt\hrule\vskip 5pt}
    02:38:20.68 & 20:06:18.7 & 17.558 & 1.068 & 0.4252 & 0.0004 & 0\\
    02:38:20.65 & 20:05:38.0 & 18.409 & 1.023 & 0.4158 & 0.0005 & 0, 1\\
    02:38:20.59 & 20:06:09.4 & 20.258 & 1.167 & 0.4140 & 0.0015 & 1\\
    02:38:20.55 & 20:06:24.0 & 20.911 & 1.050 & 0.4240 & 0.0017 & 1\\
    02:38:20.72 & 20:04:38.3 & 21.356 & 0.983 & 0.4131 & 0.0032 & 1\\
    02:38:20.36 & 20:06:20.6 & 20.083 & 1.067 & 0.4253 & 0.0005 & 2, RTT-150\\
    02:38:20.80 & 20:06:21.8 & 20.531 & 1.055 & 0.4258 & 0.0006 & 2\\
    02:38:21.10 & 20:06:23.0 & 20.895 & 1.039 & 0.4181 & 0.0015 & 2\\
    02:38:20.92 & 20:05:48.2 & 21.330 & 0.962 & 0.4095 & 0.0023 & 3\\
    02:38:21.00 & 20:05:54.0 & 20.734 & 0.956 & 0.4201 & 0.0013 & 3\\
    02:38:21.31 & 20:06:18.7 & 22.421 & 1.041 & 0.4187 & 0.0212 & 3\\
    02:38:21.86 & 20:05:45.9 & 21.677 & 1.012 & 0.4174 & 0.0020 & 4\\
    02:38:21.30 & 20:05:46.2 & 22.083 & 0.957 & 0.4229 & 0.0032 & 4\\
    02:38:18.88 & 20:05:46.1 & 19.957 & 1.028 & 0.4259 & 0.0015 & 4\\
    02:38:19.95 & 20:06:00.1 & 21.970 & 1.041 & 0.4233 & 0.0026 & 5\\
    02:38:20.75 & 20:06:07.8 & 21.251 & 0.824 & 0.4102 & 0.0015 & 5\\
    02:38:20.12 & 20:05:18.9 & 21.022 & 1.031 & 0.4186 & 0.0009 & 6\\
    02:38:21.53 & 20:05:36.1 & 20.753 & 1.056 & 0.4343 & 0.0016 & 6\\
    02:38:21.21 & 20:05:30.9 & 21.033 & 1.047 & 0.4161 & 0.0015 & 6, 7\\
    02:38:21.32 & 20:05:21.0 & 21.634 & 1.071 & 0.4188 & 0.0040 & 7\\
    02:38:20.80 & 20:06:46.3 & 21.213 & 1.066 & 0.4360 & 0.0020 & 7\\
    02:38:20.77 & 20:07:06.0 & 20.894 & 1.071 & 0.4322 & 0.0030 & 7\\
    02:38:20.55 & 20:07:21.0 & 22.128 & 0.957 & 0.4307 & 0.0289 & 7\\
    02:38:20.48 & 20:04:13.6 & 20.062 & 1.053 & 0.4244 & 0.0016 & 8\\
    02:38:20.48 & 20:04:23.8 & 20.236 & 0.959 & 0.4252 & 0.0019 & 8\\
    02:38:20.57 & 20:04:51.3 & 21.165 & 1.017 & 0.4237 & 0.0018 & 8\\
    02:38:20.60 & 20:05:45.3 & 21.986 & 1.023 & 0.4154 & 0.0024 & 8\\
    02:38:20.61 & 20:05:47.8 & 22.404 & 0.913 & 0.4158 & 0.0084 & 8\\
    02:38:23.96 & 20:05:29.3 & 22.083 & 0.957 & 0.4146 & 0.0019 & 9\\
    02:38:20.45 & 20:05:52.9 & 19.125 & 1.027 & 0.4104 & 0.0004 & RTT-150\\
    02:38:20.46 & 20:05:24.0 & 18.720 & 1.038 & 0.4213 & 0.0003 & RTT-150\\
    \noalign{\vskip 3pt\hrule\vskip 5pt}
    \label{tab:zspec}
    \end{tabular}
\end{table}

Table~\ref{tab:zspec} summarizes the spectroscopic redshift measurements for 31 galaxies in the cluster. The first four columns give the right ascensions, declinations, $r$ and $r-z$ values for the galaxies taken from the DESI photometric survey. The fifth and sixth columns present our redshift measurements and their uncertainties. For one galaxy in Table~\ref{tab:zspec}, a redshift measurement of $z_{\rm spec} = 0.4257$ \citep{desidr1} is available in the DESI spectroscopic survey and agrees within the uncertainties with our measurement of $z_{\rm spec}=0.4258$. Our measurement was obtained with the BTA telescope prior to the publication of the DESI spectroscopic catalog. The seventh column indicates the slit position labels during the observations. The label "RTT-150" marks galaxies whose redshifts were measured with the RTT-150 and reported in our previous work \citep{Peanut}. For the galaxy 02:38:20.36 20:06:20.6, the table shows the redshift measurement obtained with the BTA.

\begin{figure*}
  \centering
    \raisebox{0.2\totalheight}{\includegraphics[width=0.78\columnwidth]{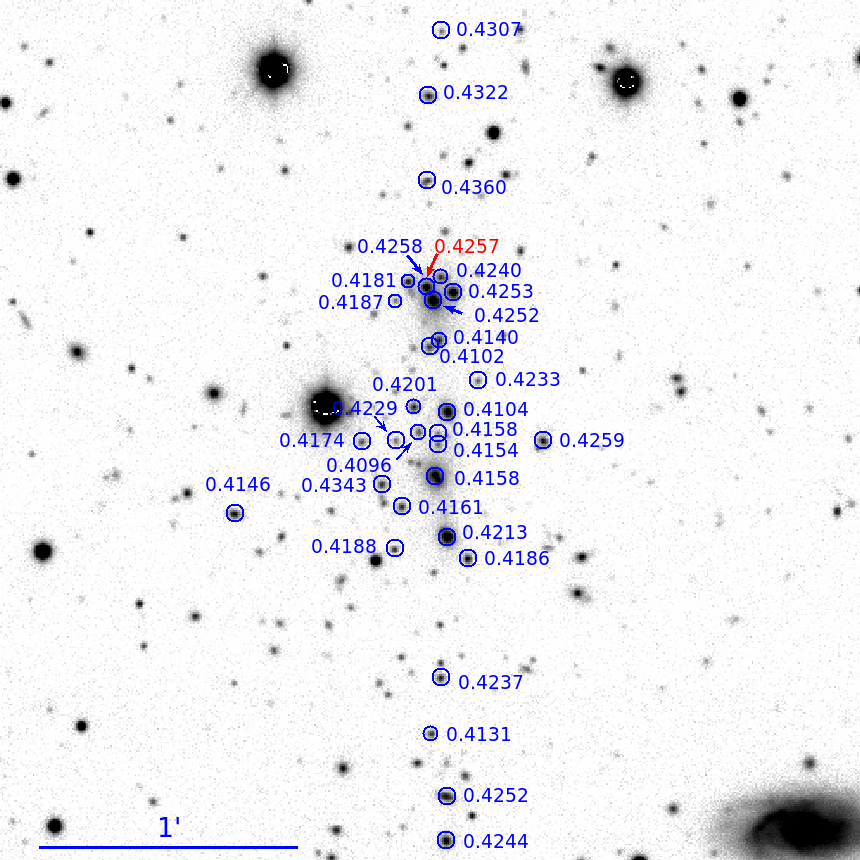}}
    \includegraphics[width=1.20\columnwidth]{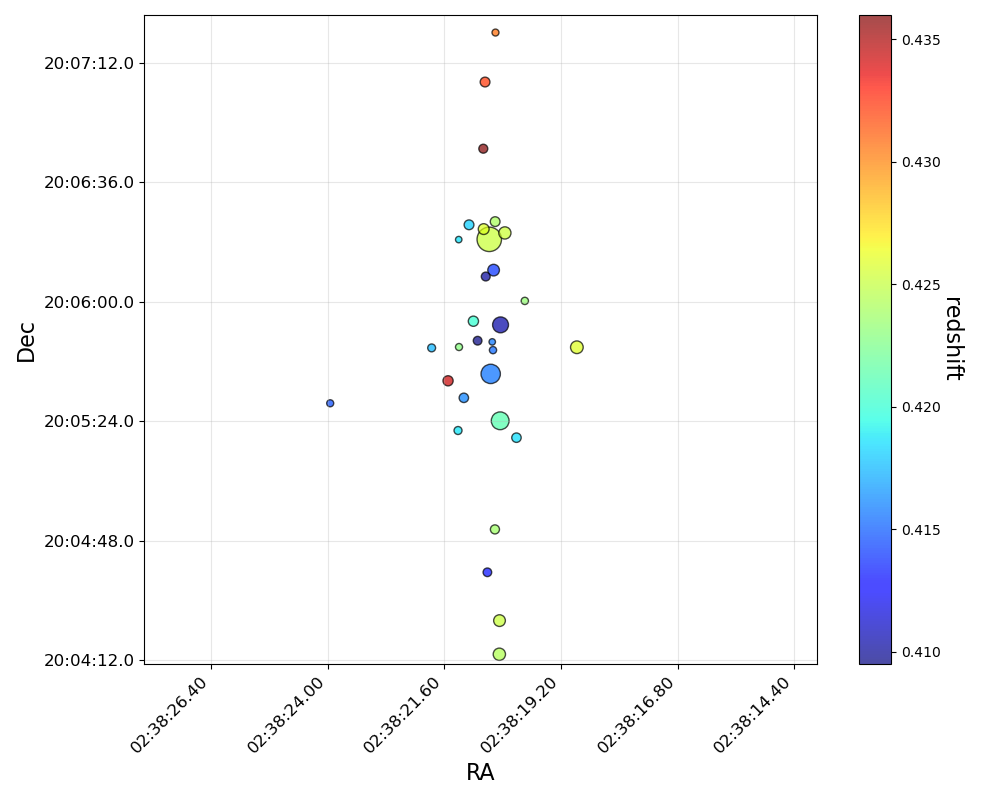}
    \caption{Measured redshifts of galaxies in the Peanut cluster. The left panel shows a DESI LIS $z$-band image. Galaxies for which we measured spectroscopic redshifts are marked in blue, with the corresponding redshift values indicated. The galaxy shown in red has a spectroscopic redshift measured by the DESI survey. The right panel shows the projected positions of the cluster galaxies; circle colors and sizes correspond to the measured redshifts and $r$-band magnitudes, respectively.}
  \label{fig:all_zsp}
\end{figure*}

\begin{figure}
  \centering
    \includegraphics[width=0.95\columnwidth]{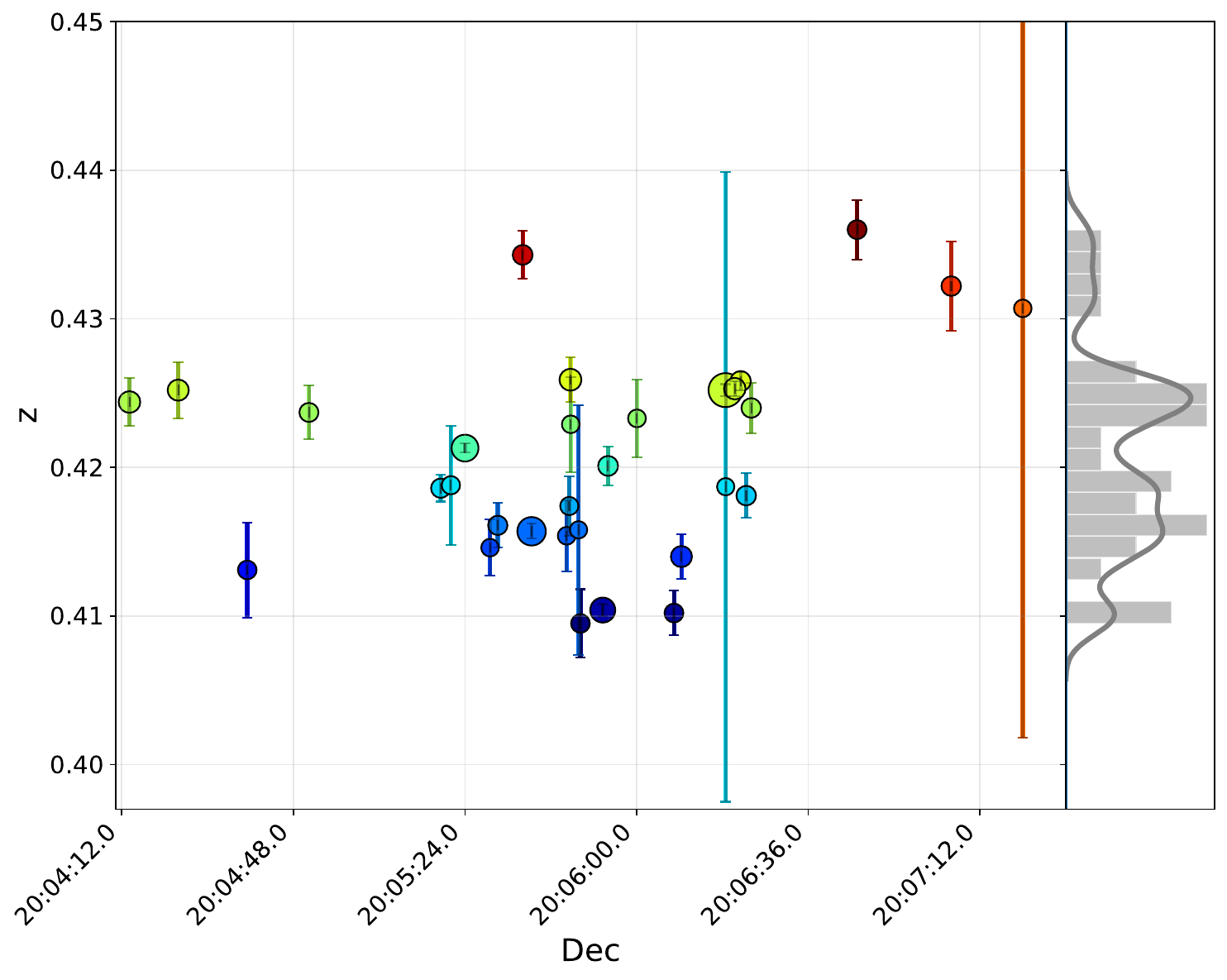}
    \caption{Spectroscopic redshifts of the Peanut cluster galaxies from Table~\ref{tab:zspec} as a function of their declination. The color of the circles represents the redshift value, using the same color scale as in Figure~\ref{fig:all_zsp} (right panel), while the symbol size is proportional to the galaxy $r$-band magnitude. The histogram shows the redshift distribution, and the gray curve represents a smoothed probability density function obtained using kernel density estimation (KDE). The distribution suggests the presence of  two components: one centered at $z \simeq 0.425$ (green–yellow circles) and another at $z \simeq 0.416$ (blue–cyan circles).}
  \label{fig:z_rainbow}
\end{figure}

Figure~\ref{fig:all_zsp} shows images with the measured galaxy redshifts indicated. The dependence of spectroscopic redshifts on galaxy declination  is shown in Figure~\ref{fig:z_rainbow}. This visualization method allows us to identify possible subclusters. The redshift histogram shows two apparent concentratoins near $z \simeq 0.425$ and $z \simeq 0.416$. However, as shown below, statistical tests indicate that the hypothesis that the observed data are drawn from a normal distribution cannot be rejected.

\begin{figure}
  \centering
    \includegraphics[width=0.95\columnwidth]{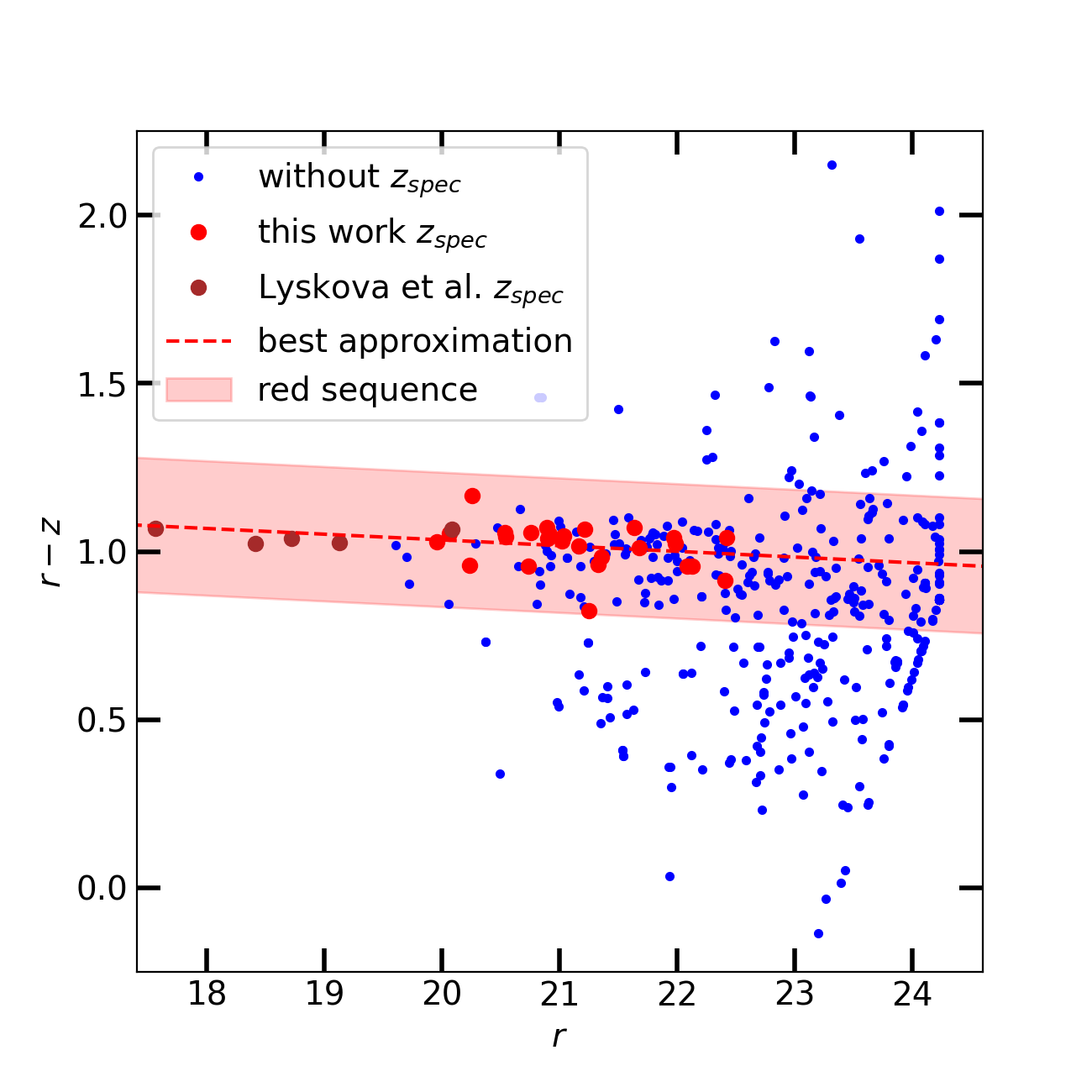}
    \caption{Color–magnitude diagram for galaxies in the cluster field. Brown circles denote the five galaxies from \cite{Peanut}, red circles mark the 26 galaxies with spectroscopic redshifts obtained in this work, and blue circles indicate galaxies without measured spectroscopic redshifts ($z_{\rm spec}$). The red dashed line shows the best-fitting relation between the $r-z$ color and the $r$-band magnitude for all galaxies with measured redshifts.}
  \label{fig:RS}
\end{figure}

Figure~\ref{fig:RS} shows the color–magnitude diagram which includes the new measurements obtained with the BTA. The refined approximation of the $r-z$ color as a function of the $r$-band magnitude for the 31 cluster galaxies yields $(r-z)_{\rm RS} = -0.014\cdot r + 1.320$. These results can be used for selecting galaxies and conducting further spectroscopic observations of the Peanut cluster.

\section{Analysis of the redshift distribution}
\label{sec:analysis}

\subsection{Substructure Identification}
\label{sec:substructure}

\begin{figure}
  \centering
    \includegraphics[width=0.95\columnwidth]{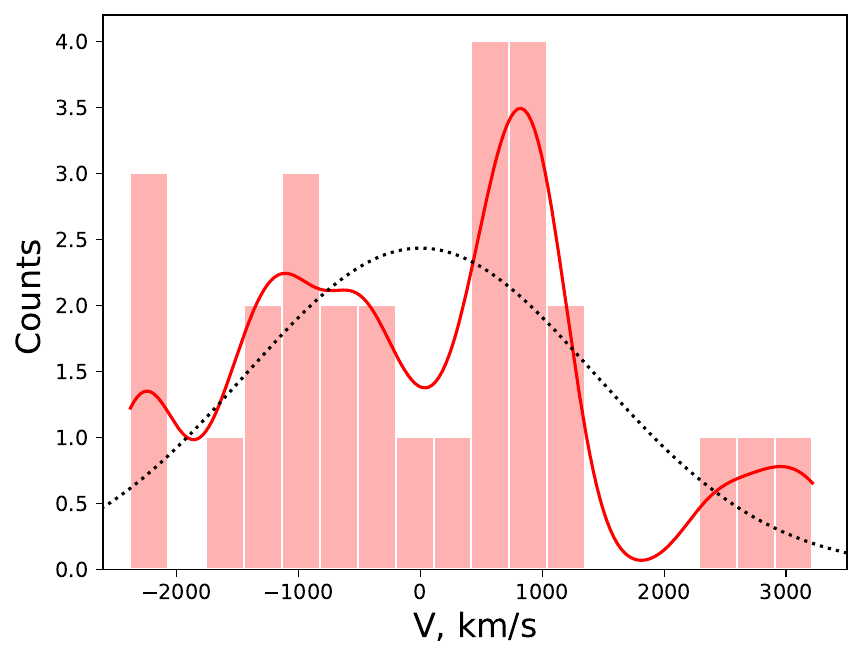}
    \caption{Distribution of line-of-sight velocities $V$ of Peanut cluster galaxies relative to the mean cluster $V$. The sample includes 28 galaxies from Table~\ref{tab:zspec} with $z_{\rm spec}^{\rm err} \leq 0.005$. The red curve shows the kernel density estimation (KDE) smoothing of the velocity histogram. The distribution suggests the possible presence of two substructures. For comparison, the black dashed line shows the best-fitting normal distribution to the data.}
  \label{fig:V_distr}
\end{figure}

\begin{figure}
  \centering
    \includegraphics[width=0.95\columnwidth]{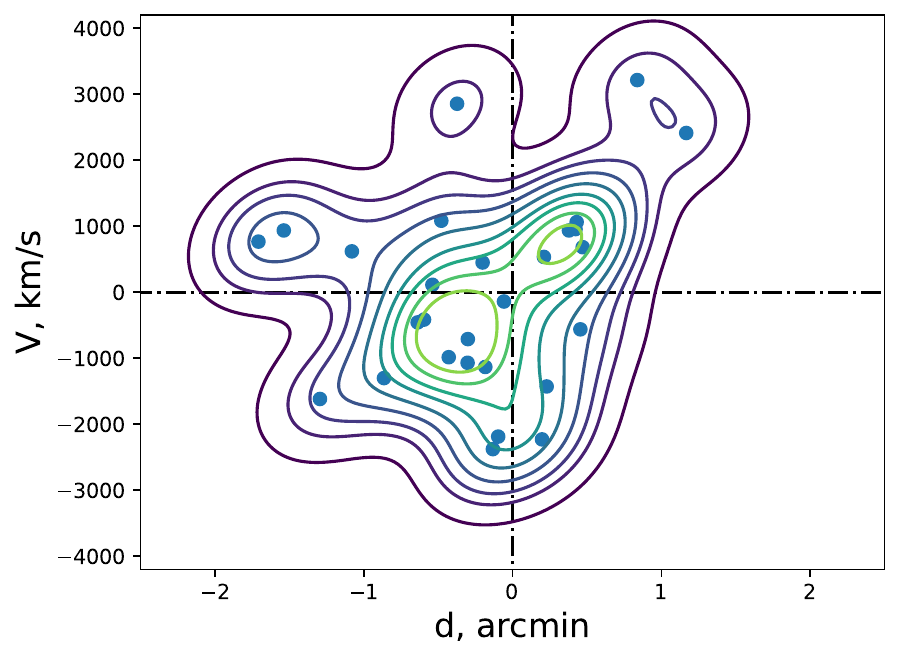}
    \caption{Visualization of the distribution of 28 Peanut cluster galaxies in the $(V, d)$ plane, where $V$ is the line-of-sight velocity relative to the mean cluster line-of-sight velocity, and $d$ is the projected distance in arcminutes from each galaxy to the X-ray cluster center. Positive and negative values of $d$ correspond to galaxies located north and south of the X-ray center, respectively. Contours show the galaxy density estimated using kernel density estimation (KDE). As in Figures~\ref{fig:z_rainbow} and~\ref{fig:V_distr}, the distribution visually suggests the presence of at least two substructures: a northern component associated with $V>0$ and $d>0$, and a southern component associated with $V<0$ and $d<0$.}
  \label{fig:density}
\end{figure}

\begin{figure*}
  \centering
    \raisebox{0.105\totalheight}{\includegraphics[width=0.922\columnwidth]{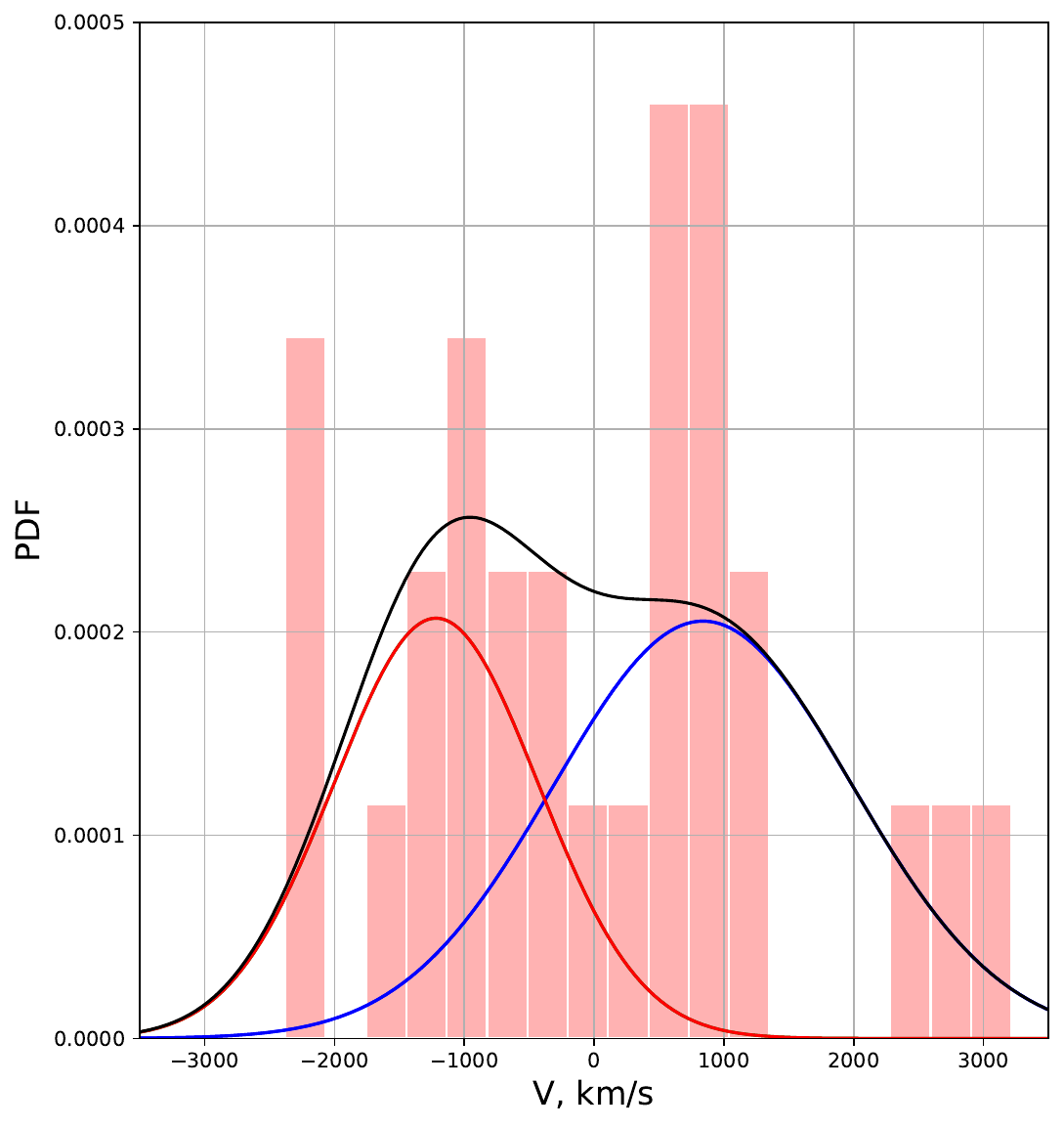}}
    \includegraphics[width=0.922\columnwidth]{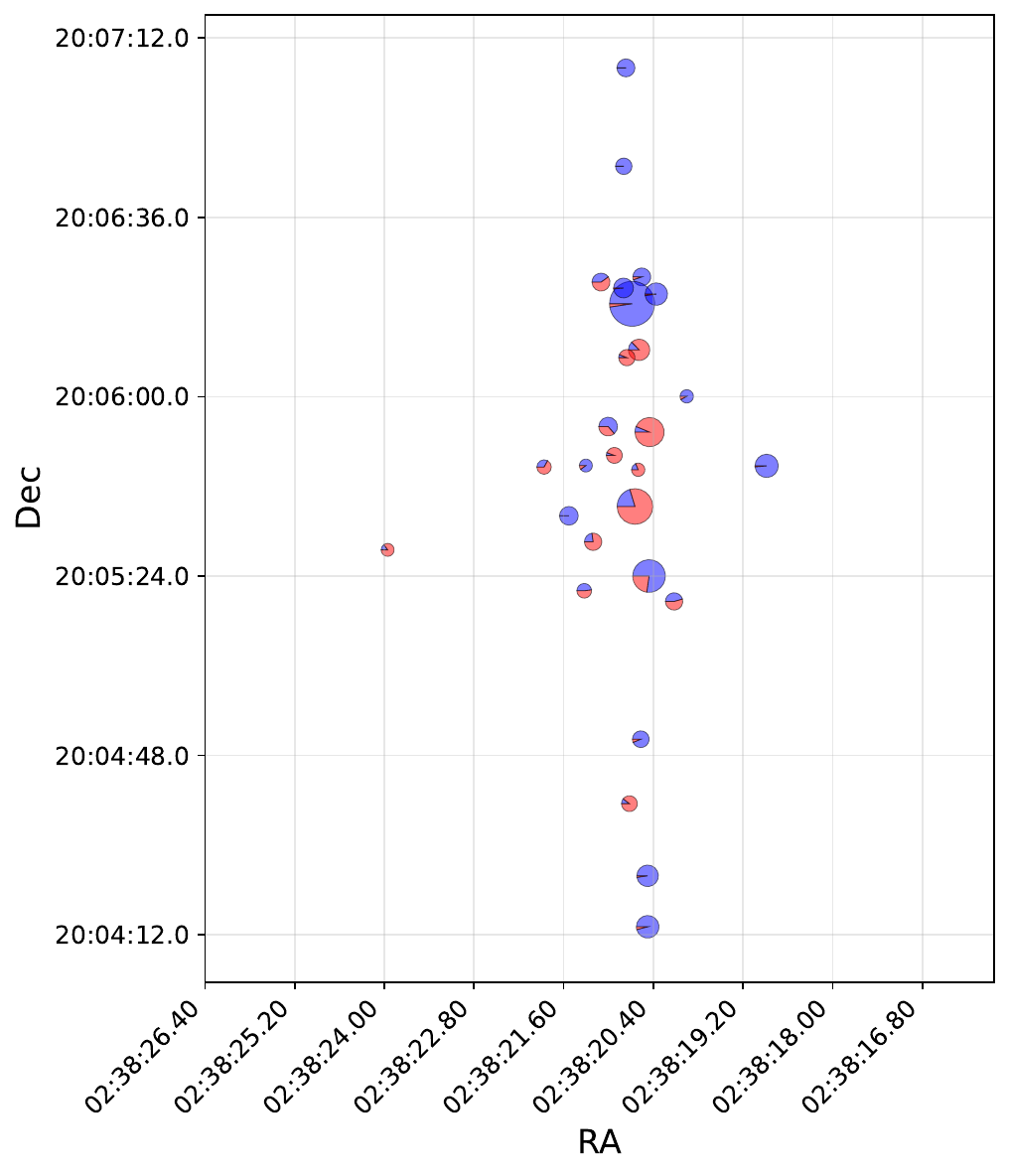}
    \caption{Normalized histogram of line-of-sight velocities of Peanut cluster galaxies for the subsample of 28 objects with $z_{\rm spec}^{\rm err} \leq 0.005$. Assuming a scenario in which two merging subclusters are present, the velocity distribution is modeled as the sum of two Gaussian components. The southern component, with a mean line-of-sight velocity of $V_1 = -1217$ km\,s$^{-1}$, is shown by the red curve, while the northern component, with $V_2 = +839$ km\,s$^{-1}$, is shown by the blue curve; their sum is indicated by the black curve. The right panel shows the projected positions of the galaxies on the sky. Circles are colored according to the probability of subcluster membership: blue for the northern subcluster and red for the southern subcluster. The symbol size is proportional to the galaxy $r$-band magnitude.}
  \label{fig:2comp}
\end{figure*}

For the subsequent analysis of the galaxy velocity distribution and spatial positions, we selected 28 galaxies with spectroscopic redshift uncertainties $z_{\rm spec}^{\rm err} \leq 0.005$, corresponding to a line-of-sight velocity uncertainty of $\sim1000$ km/s. The mean redshift of the sample is $\langle z\rangle = 0.4208$. The distribution of galaxy line-of-sight velocities, defined as $\displaystyle V = c\,(z-\langle z\rangle)/(1+\langle z\rangle)$, is shown in Figure~\ref{fig:V_distr}. The velocity dispersion, computed following the standard approach accounting for redshift measurement uncertainties \citep{Danese1980}, is $\sigma_{\rm los} = 1455 \pm 83$ km/s. All measured line-of-sight velocities lie within $\pm 2.2\,\sigma_{\rm los}$. In addition, the targets were selected among the brightest galaxies in the central cluster region with $r \lesssim 22$. Therefore, it is unlikely that the analyzed sample contains foreground or background galaxies projected onto the cluster field \citep[e.g.,][]{1977ApJ...214..347Y}.

As discussed above, a comparison of the X-ray and optical images of the Peanut cluster suggests that a merger of two subclusters is likely being observed \citep{Peanut}. 
Possible substructure in the distribution of line-of-sight velocities and galaxy positions is also hinted at by the analysis of Figures~\ref{fig:z_rainbow}, \ref{fig:V_distr}, and \ref{fig:density}. The latter figure shows the distribution of galaxies in the $(V,d)$ plane, where $V$ is the line-of-sight velocity relative to the mean cluster velocity and $d$ is the projected distance from each galaxy to the X-ray center, defined to be positive for galaxies located north of the X-ray center and negative for those located south. The contours represent isodensity levels of the galaxy distribution in the $(V,d)$ plane derived using kernel density estimation (KDE).

If the Peanut cluster is interpreted as a merger of two subclusters and the line-of-sight velocity distribution of each component is assumed to follow a normal distribution, then the mean line-of-sight velocity and velocity dispersion of each subcluster can be estimated. An illustration of the two-Gaussian fit to the observed velocity distribution is shown in Figure~\ref{fig:2comp}. The southern subcluster is characterized by a mean line-of-sight velocity of $V_S \approx -1200$ km/s and a dispersion of $\sigma_S \approx 800$ km/s, while the northern subcluster has $V_N \approx +800$ km/s and $\sigma_N \approx 1150$ km/s. Within this framework, the inferred relative merger velocity of the subclusters is $\simeq 2000$ km/s, consistent with estimates from  \cite{Peanut}.
The right panel of Figure~\ref{fig:2comp} shows the projected positions of the galaxies on the sky. The circles are colored red and blue according to the relative contributions of the southern and northern components to the total velocity distribution shown in the left panel of Figure~\ref{fig:2comp}, respectively.

Using the scaling relation between velocity dispersion and halo mass from \cite{2013MNRAS.430.2638M}, the derived velocity dispersions formally correspond to $M_{200}^{S} \approx 4\times10^{14}\,M_\odot$ and $M_{200}^{N} \approx 10^{15}\,M_\odot$ for the southern and northern subclusters, respectively. However, several caveats should be noted. First, dynamical mass estimates for merging systems are known to be biased high on average (e.g., \citealt{2018MNRAS.475..853O, 2019AAS...23343804T}). 

Second, while the two-Gaussian model provides an adequate description of the data, it yields no statistically significant improvement over a single normal distribution. 
Indeed, the relative improvement of the two-component fit, calculated via the likelihood ratio $2{\ln\mathcal{L}_2/\mathcal{L}_1}$ is $\sim 1-2$ for several local likelihood maxima (one of them is shown in Figure~\ref{fig:2comp}) for 2-3 additional degrees of freedom (depending on whether the width of the second Gaussian is a free parameter or not). The corresponding p-values are $\sim0.3-0.6$, precluding any strong conclusion on the presence of two components. The problem, of course, stems from the limited number of galaxies in the sample.

To assess the statistical significance of the visually identified substructures, we performed normality tests on the distribution of the measured galaxy redshifts. Given the relatively small sample size, we applied the Shapiro–Wilk test \citep{shapiro} and D’Agostino’s $K^2$ test \citep{dagostino}. The Shapiro–Wilk test yields a statistic of 0.9616 with a $p$-value of 0.3809, while D’Agostino’s test gives a statistic of 0.7293 with a $p$-value of 0.6945. These results do not allow us to reject the null hypothesis that the galaxy redshift distribution in the Peanut cluster is consistent with a normal distribution.

\cite{1988AJ.....95..985D} proposed a statistical test designed to identify local deviations in galaxy kinematics from the global dynamical properties of a cluster. Application of this test to our data does not reveal statistically significant evidence for a presence of gravitationally bound substructures in the observed distribution of line-of-sight velocities and galaxy positions. However, it should be noted that a non-detection in the Dressler–Schectman test cannot be interpreted as evidence that the cluster is dynamically relaxed or free of substructure \citep{2012MNRAS.419.1017C}.

\subsection{Velocity Dispersion and the Cluster Mass}
\label{sec:dispersion}

As shown by \cite{2020A&A...641A..41F}, estimates of galaxy cluster velocity dispersion may be biased when based on a small number of spectroscopic measurements ($N_{\rm gal}<75$). Using mock galaxy clusters drawn from cosmological simulations, these authors derived a correction that explicitly accounts for the number of available measurements $N_{\rm gal}$ and enables an approximately unbiased estimate of the velocity dispersion. Applying this correction to the Peanut cluster yields $\sigma_{\rm los}^{\rm cor} \simeq 1470$ km/s, corresponding to $M_{200} \simeq 2\times10^{15}\,M_{\odot}$, where the cluster mass is inferred from the mass–velocity dispersion scaling relation of \cite{2013MNRAS.430.2638M} following the prescription of \cite{2020A&A...641A..41F}. As noted above, dynamical mass estimates for merging systems are generally biased high on average (e.g., \citealt{2018MNRAS.475..853O, 2019AAS...23343804T}). However, \citealt{2018MNRAS.475..853O} show that for the most massive galaxy clusters with $M_{200}\sim10^{15}\,M_{\odot}$, such as the Peanut cluster, this bias becomes negligible. 

Using the halo mass function of \cite{2008ApJ...688..709T} within a $\Lambda$CDM cosmology, the expected number of halos above a given mass threshold at high redshift can be estimated. At the high-mass end, the abundance of clusters  is highly sensitive to the adopted cosmological parameters ($\Omega_m$, $\sigma_8$, $n_s$), with the strongest dependence on $\sigma_8$. Adopting the Planck  \citep{2020A&A...641A...6P} or the WMAP \citep{2013ApJS..208...20B} cosmological parameters, the predicted number of halos with $M_{200} \geq 2 \times 10^{15}\, M_\odot$ at $z \geq 0.42$ is $\simeq 15$--$17$ over the full sky. This prediction also depends on the specific halo mass function adopted. Using the \cite{2008ApJ...688..709T} or the \cite{2016MNRAS.456.2361B} mass functions yields a total of  $\sim15$ such halos, whereas for the \cite{2016MNRAS.456.2486D} HMF predicts a number $\simeq 50\%$ higher. In the ACT DR6 catalog \citep{2026OJAp....955863A}, eight such clusters with $M_{200} \geq 2 \times 10^{15}\, M_\odot$ at $z \geq 0.42$ are identified within the $16{,}293\,\mathrm{deg}^2$ survey footprint. Extrapolating to the full sky yields $\sim 20 \pm 7$ such clusters, i.e. in good agreement with the theoretical predictions. The Peanut cluster, therefore, appears to be a rare system and a particularly promising target for a detailed study, comparable to such notable  merging clusters such as the Bullet Cluster and El Gordo.

\section{Conclusion}

We have presented the results of spectroscopic redshift measurements for 31 galaxies in the Peanut cluster, including 26 new measurements obtained with the BTA telescope.
The line-of-sight velocity distribution of galaxies in the Peanut cluster reveals a possible presence of two subclusters, which is consistent with the conclusions from our previous work \citep{Peanut} based on the analysis of X-ray observations in comparison to the optical image. The line-of-sight component of the merger velocity between the northern and southern subclusters is $\sim 2000$ km/s. However, statistical tests and the Dressler-Schectman test do not allow us to reject the hypothesis that the cluster velocity distribution is consistent with a single normal distribution and therefore provide no statistically significant evidence for the presence of gravitationally bound substructures. Assuming a normal velocity distribution in the Peanut cluster, we obtained estimates of the velocity dispersion of $\sigma_{los} = 1455 \pm 83$~km/s and mass of $M_{200} \simeq 2\times 10^{15}$ $M_{\odot}$ based on the mass–velocity dispersion scaling relation. This high mass makes the Peanut cluster a rare and particularly interesting system for detailed study, comparable to extreme clusters such as the Bullet Cluster and El Gordo. To reach a definitive conclusion regarding the presence of substructures in the cluster, additional spectroscopic redshift measurements for several tens of galaxies in the Peanut cluster are required. For the purpose of selecting candidate galaxies, the measured $r-z$ color of the cluster red sequence, $(r-z)_{RS} = -0.014\cdot r + 1.320$, can be used.

\section*{Acknowledgements}

The observations at the 6-m BTA telescope of the Special Astrophysical Observatory of the Russian Academy of Sciences (SAO RAS) are supported of the Ministry of Science and Higher Education of the Russian Federation. The renovation of telescope equipment is currently provided within the national project "Science and universities." The work of AA, SD, SK, AM, and RU was carried out within the state assignment of the SAO RAS, approved by the Ministry of Science and Higher Education of the Russian Federation.

This work made use of data from the DESI Legacy Imaging Surveys. The DESI Legacy Imaging Surveys consist of three individual projects: the Dark Energy Camera Legacy Survey (DECaLS), the Beijing-Arizona Sky Survey (BASS), and the Mayall z-band Legacy Survey (MzLS). DECaLS, BASS and MzLS together include data obtained, respectively, at the Blanco telescope, Cerro Tololo Inter-American Observatory, NSF's NOIRLab; the Bok telescope, Steward Observatory, University of Arizona; and the Mayall telescope, Kitt Peak National Observatory, NSF's NOIRLab. The Legacy Surveys project is honored to be permitted to conduct astronomical research on Iolkam Du'ag (Kitt Peak), a mountain with particular significance to the Tohono O'odham Nation.

The authors are grateful to T\"UB\.ITAK, IKI, KFU, and the Academy of Sciences of Tatarstan for partial support in using the RTT-150 (Russian-Turkish 1.5-m telescope in Antalya).

The work of IB, IKh, MS was supported by subsidy 30000P.\-16.\-1.\-OH17AA81027, allocated to TAS IAS to fulfill the state assignment in the field of scientific activity.

\clearpage

\bibliographystyle{elsarticle-harv}
%\bibliographystyle{mnras}
%\bibliography{refs}

%=================================

\end{document}